\documentstyle[12pt,epsf,rotating]{article}

\def\beq{\begin{equation}}   \def\eeq{\end{equation}}

\newcommand{\gsim}{\lower.7ex\hbox{$\;\stackrel{\textstyle>}{\sim}\;$}}
\newcommand{\lsim}{\lower.7ex\hbox{$\;\stackrel{\textstyle<}{\sim}\;$}}

\newcommand{\ra}{\rightarrow}

\newcommand{\eq}[1]{Eq.\hspace*{.1em}(\ref{#1})}

\renewcommand{\Im}{\mbox {Im}\:}

\newcommand{\GeV}{\,\mbox{GeV}}
\newcommand{\MeV}{\,\mbox{MeV}}

\begin{document}

\def\lsim{\mathrel{\rlap{\lower3pt\hbox{\hskip0pt$\sim$}}
    \raise1pt\hbox{$<$}}}         
\def\gsim{\mathrel{\rlap{\lower4pt\hbox{\hskip1pt$\sim$}}
    \raise1pt\hbox{$>$}}}         

\begin{titlepage}
\renewcommand{\thefootnote}{\fnsymbol{footnote}}

\begin{flushright}
CERN-TH/96-252\\
TPI-MINN-15/96-T\\
UMN-TH-1509/96\\
UND-HEP-96-BIG\hspace*{.15em}$05$\\
hep-ph/9610515\\

\end{flushright}

\vspace{0.5cm}

\begin{center}
\baselineskip25pt

{\Large\bf Chiral Symmetry Breaking,  Duality in the
$\bar Q q$ Channel and $b\ra \bar c c s$ Decays}

\end{center}

\vspace{0.5cm}

\begin{center}
\baselineskip12pt

\def\thefootnote{\fnsymbol{footnote}}

{\large B. Blok$^{1,3}$, M.~Shifman$^{2,3}$} and {\large N.~Uraltsev$^{3,4,5}$}

\vspace{0.5cm}

{\it $^1$ Department of Physics, Israel Institute of Technology,
Technion, Haifa 32000, Israel$^{\:\dagger}$ \\[0.3cm]

$^2$  Theoretical Physics Institute, University of Minnesota, Minneapolis,
MN 54555 USA$^\dagger$ \\[0.5cm]

$^3$ TH Division, CERN,  CH-1211 Geneva 23, Switzerland\\[0.25cm]

$^4$ Department of Physics, University of Notre Dame du Lac, Notre Dame,
IN 46556, U.S.A.\\[0.25cm]

$^5$ St.Petersburg  Nuclear Physics Institute, Gatchina, St.~Petersburg,
188350 Russia$^{\,\dagger}$
}

\vspace{1cm}

{\large\bf Abstract} \vspace*{.25cm}
\end{center}

We address  the issue of the quark-hadron duality
in the spectral densities induced by the heavy-light quark currents
$\bar Q q$. In the limit $m_Q\ra\infty,\;\; m_q\ra 0$
we observe an enhancement of the physical spectral density compared
to the quark one in the scalar and axial channels, due to the
Goldstone meson contributions.
This may imply that the scale where duality sets in  is higher in these 
channels
 than in the vector (pseudoscalar) case.
Implications for the nonleptonic decays of
$B$ mesons (the $b\rightarrow \bar ccs$ transition)
are considered. We discuss in detail the decay pattern and obtain
an independent estimate of the ``wrong" sign $D$ yield.

\hfill

\begin{flushleft}
CERN--TH/96--252\\

September 1996

\vspace{0.25cm}

\rule{2.4in}{.25mm} \\
$^\dagger$ Permanent address.

\end{flushleft}

\end{titlepage}

\section{Introduction}

 A large and rapidly growing number of applications of the heavy quark
theory requires
predictions in the Minkowski domain. The inclusive width of
 the heavy flavour hadrons is one the best known examples. All
predictions of this type are based on the so-called QCD duality
(for a recent discussion see, e.g. Ref. \cite{1}).

The issue of duality is as old as QCD itself. Because of its complexity it
was virtually neglected for a long time, basically after
the classical paper of Poggio {\em et al.} \cite{PQW}.  Recently the interest to
this
question was revived by the necessity of having QCD predictions valid
up to  per cent accuracy in several problems of practical importance.
The first attempts at quantifying possible deviations from duality
were presented  in Refs. \cite{2,3}.

It is not easy  to estimate the scale of the duality violations from
fundamental QCD. Therefore, it is of paramount importance to gain experience 
in
various particular problems.
The present work investigates duality violations in an applied aspect.
We will be mostly concerned with
the spectral density in the two-point function induced by the
heavy-light currents $\bar Q \Gamma q$, where $Q$ and $q$ are the light and
heavy
quarks, respectively.
It turns out that
in the limit $m_Q\rightarrow\infty$, $m_q\rightarrow 0$ the
spectral density in the scalar (axial) channel is very peculiar:
it starts {\em below} the position of the ground-state resonance,
and is enhanced by the Goldstone meson contribution. Qualitatively the 
emerging picture is quite different from
that
taking place in the ``conventional" vector (pseudoscalar) channel, where the
ground state resonance is physically the lowest-lying state. A key role in this
phenomenon belongs to the spontaneous breaking of the chiral symmetry
for the light quarks.

The formulation of the problem to be discussed below is, in a sense,
complementary to Ref. \cite{2}. If in the latter work
one approaches from the high-energy side (i.e. one considers 
the oscillation and asymptotic zones, in the nomenclature of Ref.~\cite{2},
see Sect.~5.2), where duality is defined point-by-point; here we mostly deal
with the resonance zone. Correspondingly, by duality we mean here that
certain finite-range integrals over the hadronic spectral density are equal
to the same integrals over the quark spectral density.  The equality (or,
rather, deviations from it) is checked by saturating the hadronic
contribution by exclusive modes (cf. Ref. \cite{3}).  The size of an
integration interval large enough for duality to take place characterizes
the scale of structures and disturbances due to low-lying resonances. 

In the second part of the paper, as a practical application we consider  a 
particular physical process, namely
$b\rightarrow c\bar c s$. Due to a relatively small energy release in
this transition it is a natural suspect
\cite{buv,4,5} for duality violations, which might be relevant in the 
so-called ``semileptonic branching ratio versus $n_c$ problem" \cite{6}. The
model of Ref. \cite{2} estimates possible violations of duality in the
$b\rightarrow \bar c cs$ only at the per cent level. If so, they can be
 neglected, and the duality-based predictions \cite{BBSL,volos} 
should be valid.
The above model is admittedly crude, however, and independent estimates 
are badly needed.
Since we observe a strong enhancement of the spectral density
in the axial $\bar c s$ channel, associated with the soft kaon contribution,
a corresponding enhancement in the $\bar B\rightarrow D^{*}\bar D^{*}\bar 
K$ could be naturally expected. {\em A priori}, it could then be duality
violating.

We suggest to use factorization, and the so-called generalized small velocity 
(GSV) limit in the $b\rightarrow c\bar c s$ transition. Both approximations are
described at length below; here we just note that using them opens
all decays of the type $\bar B \ra D \bar D +\,\,\, S$-wave kaons (kaons and 
pions) for applications of the soft pion technique \cite{VZ}. The most economic 
way to implement the idea is by using the effective Lagrangian approach, 
which  combines the chiral and heavy-quark symmetries \cite{16,A,B}.

Surprisingly, our estimates of the decay modes $\bar B \ra D^{(*)} \bar D^{(*)} 
\bar K$, where the kaon is treated as a soft Goldstone boson,
give no indications on significant duality violations. We hasten to add, though, 
that the accuracy of the above estimates is not high.

The mechanism we discuss does not differentiate  noticeably 
between the $B$ meson and $\Lambda_b$ baryon decays.
Thus, it adds nothing new to the problem
of the $\Lambda_b$  lifetime. However, the $1/N_c$-motivated factorization 
used in the $B$ decays  may no longer be applicable when a heavy baryon is
produced in the final state of $\Lambda_b$ decays. This may provide a 
mechanism
for the observed lifetime difference, but we will not go into this issue here.

An interesting lesson follows for the lattice analysis
of the heavy-light systems in the scalar and axial channels,
where the quenched approximation is expected to give results
different from those with light dynamical quarks.

 The paper  consists of two distinct parts. In the first part
we discuss general features of the spectral densities
corresponding to the two-point functions induced by
$\bar Q q$ in the limit $m_Q\ra\infty,\,\,\, m_q\ra 0$.
(Sect.~2). Implications for the transition $b\ra \bar c c s$
are considered in the second part (Sects.~3 and 4). Our conclusions are 
summarized in Sect.~5.

 \section{Spectral densities and duality in the scalar and pseudoscalar
 $\bar Q q$ channels}

 As a preparatory step let us consider the spectral densities for the
two-point functions induced by
the current $\bar Q\gamma_\mu (1-\gamma_5)q$. Since our consideration
at this stage focuses on qualitative aspects, it is convenient to work in the
limit where the mass of the heavy quark $m_Q\rightarrow \infty$ while that
of the light quark $m_q\rightarrow 0$. In the chiral limit all mesons
($\pi, K, \eta )$ belonging to the Goldstone octet are massless. The effects due
to  the finite quark masses will be incorporated later,
when we pass to $b\ra \bar c c s$. The vector and pseudoscalar
 $\bar Q q$ mesons are degenerate. The same is valid for the
 scalar and axial ones. The splitting between the first and the
second multiplets is
of order $\Lambda_{\rm QCD}$ (practically, about 500 MeV).

 The first analysis of the two-point functions
$$
\Pi_{\mu\nu}^{(V)}= i\int d^4x \, {\rm e}^{iqx}\, \langle 0\vert V_\nu (x)
V^\dagger_\mu (0)\vert
0\rangle =
\Pi\, g_{\mu\nu}+...\,  ,
$$
$$
\Pi_{\mu\nu}^{(A)}= i\int d^4x\, {\rm e}^{iqx}\, \langle 0\vert A_\mu (x),
A^\dagger_\nu (0)\vert 0\rangle
= \tilde\Pi \, g_{\mu\nu}+...\, ,
$$
$$
\Pi^{(S)}=i\int d^4x \, {\rm e}^{iqx}\, \langle 0 \vert S(x), S^\dagger (0)\vert
0\rangle \, ,
$$
\beq
\Pi^{(P)}= i\int d^4x \, {\rm e}^{iqx}\, \langle 0 \vert P(x),P^\dagger (0)\vert
0\rangle\, ,
\label{1}
\eeq
dates back to the early days of QCD \cite{8}. The currents $V_\mu ,\,\,
A_\mu ,\,\, S$ and $P$
are defined as 
\begin{equation}
 V_\mu=\bar Q\gamma_\mu q,\,\,\,
A_\mu=\bar Q\gamma_\mu\gamma_5q,\,\,\,
S=\bar Qq,\,\,\,
P=\bar Qi\gamma_5q \, .
\label{2}
\end{equation}
Although the weak currents are $V-A$, the scalar and pseudoscalar
two-point functions appear as their longitudinal parts,
$$
S=\frac{-i}{m_Q-m_q}\, \partial _\mu V_\mu\;\; , \;\;\;\;\;
P=\frac{1}{m_Q+m_q}\, \partial _\mu A_\mu\;\; .
$$

Already in Ref. \cite{8} it was noted that the quark condensate correction is
enhanced in the heavy-light quark case compared to, say, the classical
$\rho$-meson sum rule \cite{9}, producing stronger disturbances of the
spectral
density at small energies and, thus, leading to stronger deviations from
duality \footnote{We stress again that we focus now on the resonance zone,
where a typical scale
is set by low-dimension condensates, unlike Ref. \cite{2} where
the emphasis was on the oscillation and asymptotic zones, and 
on the
modelling of
the impact of high-dimension condensates.}. First of all, its dimension in
the former case is 3, while in  the latter case it is 6. Second, the
numerical coefficient is larger. Further studies \cite{10,11,Dai} confirmed this
observation in a more quantitative way. Actually, Ref. \cite{10} was the
first to introduce the heavy quark limit. It was noted there that in this
limit the spectral densities in the (transverse)  vector and
pseudoscalar channels are degenerate.  The same is valid for the scalar and 
(transverse)
axial channels. For  this reason we will limit ourselves in this section
only to $S$ and $P$ channels.

Even more remarkable is the fact \cite{10} that using the standard
``first resonance plus parton-like continuum" model of the spectral density
produces an abnormally large residue  of the ground-state resonance in the
scalar (axial) channels, an order of magnitude larger than that in the
pseudoscalar (vector) channels.  This observation,   which went unnoticed,
must be viewed as a hint that something unusual takes place in the scalar
channel. Now we are ready to explain this anomaly.

As a matter of
fact, the standard model of the spectral density mentioned above does not
work in the scalar (axial) channels because of an  enhancement
 at low energies due to the
  contribution of the $D\pi$
 intermediate state. (The mesons of the type $D$ and $D^*$ built from
$Q\bar q$ and degenerate in the limit
$m_Q\rightarrow \infty, m_q\rightarrow 0$ will be generically referred
to as $D$'s. The massless pseudoscalar mesons will be generically referred to
as $\pi$'s).

In the free-quark approximation the four spectral densities
(for $\Pi^{(V,P)}$
and $\Pi^{(S,A)}$) are all identical and equal to
\begin{equation}
{\rm Im}\, \Pi=\frac{N_c}{2\pi}\, \varepsilon^2 \; ;
\label{3}
\end{equation}
only  chirally odd condensate corrections distinguish between the
first and the second pairs of  above currents.
Here $\varepsilon$ is the excitation energy measured from the quark
threshold (i.e. from the heavy quark mass $m_Q$).  Logarithmic factors due
to anomalous dimensions and $\alpha_s$ corrections are neglected in Eq.
(\ref{3}). Although they play a role numerically, see e.g.  Ref. \cite{11},
qualitatively they are not important. Equation  (\ref{3}) will set a natural
scale for the spectral density in the heavy-light channels. Its important
feature is a strong suppression at low energies, as $\varepsilon^2$.

Now, let us turn to  the physical (hadron-saturated) spectral densities.
In the pseudoscalar channel we encounter a more or less familiar picture.
The physical spectral density starts from the ground-state pseudoscalar
$D$, then there is a gap, and then multiparticle intermediate states
(continuum) add up into a curve that is expected to be  relatively
close to the quark one. The continuum actually is a sum over broad 
resonances, the radial excitations of $D$.

The first resonance in the pseudoscalar (vector)
channel is situated at $\varepsilon =\bar \Lambda$ where $\bar\Lambda 
=
M_D-m_c$; the continuum threshold is at $\varepsilon =
\varepsilon_c\sim 2 \bar \Lambda$, see Fig.~1. The onset of ``continuum" 
approximately coincides with the position of the first radial excitation in the 
given channel. The parameter $\bar\Lambda$ above is a basic 
parameter of the heavy quark  theory \cite{Luke}. 
Numerically $\bar\Lambda \sim 500$ MeV \cite{NS,Baga,MV}.
(We 
recall
that  although we keep using the name ``$D$ mesons"
so far we work  in the limit $m_Q\ra\infty$.)

The scalar (axial) resonances we are interested in are $P$-wave in the 
language of the
naive quark model  and $1/2^+$   in the language of HQET.  Due to the 
chiral symmetry breaking they lie higher.\footnote{The scalar 
resonance will be referred to as $D_0^*$. The axial
$1/2^+$ resonances are sometimes denoted as $D_1'$, to 
distinguish them from the axial $3/2^+$ resonances denoted as $D_1$.
Since in the present paper we will discuss only the axial $1/2^+$ 
resonances, the prime will be omitted. The primes are reserved for
radial excitations.} (For a concise review of the higher $\bar Q q$ states
see Ref. \cite{Adam}.) The heavy quark symmetry
implies that the scalar $D^*_0$ and the axial $D_1$ are degenerate; one
expects both  $D^*_0$ and $D_1$  at $\varepsilon\sim$ 2$\bar\Lambda$.
What is remarkable is that the scalar (transverse axial) spectral density is 
non-vanishing
and large {\em below } $D^*_0$ ($D_1$),  in the interval from
$\bar\Lambda$ to $\sim 2 \bar\Lambda$, since  it
 receives an enhanced contribution from the states $D\pi$
(in the scalar channel) and $D^*\pi$ (in the axial channel). The pions are
strongly
coupled to the $\bar Q q$ current, in the $S$ wave.

\begin{figure}
\vspace{5.0cm}
\includegraphics{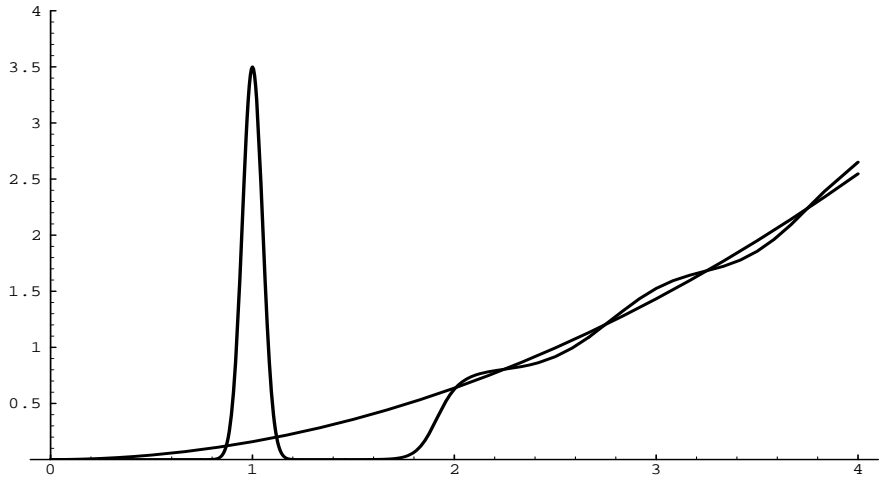}
\caption{A sketch of the spectral density
versus $\varepsilon /\bar \Lambda$
in the pseudoscalar channel. To set the scale we show also the free quark
spectral density represented by the parabola.
}
\end{figure}
 
\begin{figure}
\vspace{4.0cm}
\includegraphics{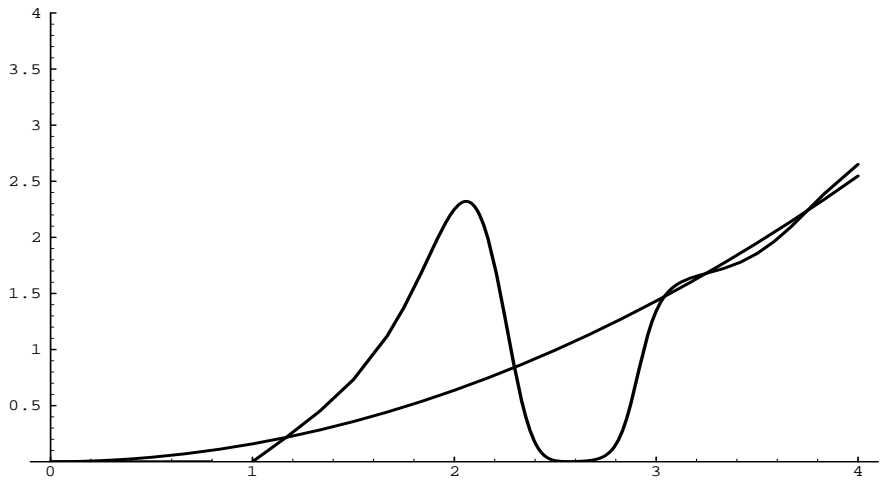}
\caption{A sketch of the spectral density
 versus $\varepsilon /\bar \Lambda$
in the scalar channel.
}
\end{figure}
 
Consider for definiteness $\Pi^{(S)}$.  In the  low-energy limit
the pion is soft and  can be reduced by using the soft-pion technique. Then,
\begin{equation}
\langle 0\vert \bar Qq\vert D \pi \rangle_{\vec k\ra 0} =
\frac{1}{f_\pi}\langle
0\vert\bar Qi\gamma_5q\vert D\rangle =
-\frac{f_Q}{f_\pi}\, m_Q\, .
\label{4}
\end{equation}
Here $f_Q$ is defined as
$$
\langle 0\vert \bar Q\gamma_\mu\gamma_5q\vert D\rangle =if_Qp_\mu\, ,
$$
and we do not differentiate between $M_D$ and $m_Q$ in the limit
$m_Q\rightarrow\infty$; $\vec k$ is the pion momentum. The spectral
density takes the form
\beq
{\rm Im}\, \Pi^{(S)}=\frac{1}{2}\vert\langle 0\vert\bar Qq\vert D\pi\rangle
\vert^2\times
\mbox{phase space} = \frac{f^2_Qm_Q}{f^2_\pi}\frac{1}{8\pi}\left( N_f-
\frac{1}{N}_f\right)\, (\varepsilon-\overline{\Lambda})\; ,
\label{5}
\eeq
where $N_f=3$ is the number of  massless flavours ($N^2_f-1$ is the 
number
of  Goldstone mesons) and $\varepsilon-\overline{\Lambda}$ is
the energy measured from the position of the ground state $D$.

Comparing Eqs. (\ref{3}) and (\ref{5})
 reveals the remarkable enhancement
of the spectral density at low energies that was mentioned above.
First, the dependence on $\varepsilon$ is parametrically different.
Due to the $S$-wave nature of the matrix element
 $\langle 0\vert S\vert D\pi\rangle $ the $D\pi$ contribution to the spectral
density
 vanishes as $O(\varepsilon )$ while the free quark expression (\ref{3}) is
O($\varepsilon^2)$. Second, as is well known, the static coupling $f_Q$ is
rather large numerically \cite{11,12}. The combination $f^2_Qm_Q$ stays
constant in the limit $m_Q\rightarrow\infty$ \cite{10}, modulo the hybrid
logarithms \cite{13}, and this constant is approximately 0.2 to 0.25 GeV$^3$
\cite{11,12}. Treating  Eq. (\ref{5})
literally we get  that the $D\pi$ contribution
to the spectral density at $\varepsilon = 2\bar\Lambda$ exceeds the free
quark expression by a factor of $\sim 1.5$ (Fig. 2).
The integral over the $D\pi$ spectral density up to
$\varepsilon = 2\bar\Lambda$ is approximately $1.2$ times the integral over
the free quark spectral density from 0 to $\varepsilon=2\bar\Lambda$.

Of course, at a certain energy, the soft-pion
result (\ref{4}) becomes invalid -- the momentum-dependent part of the
interaction will cut off the amplitude. The cut off presumably occurs
near  the position of the resonance $D^*_0$. The excess of the spectral
density below $D^*_0$ and at $D^*_0$ has to be compensated by an extended 
gap
immediately
after $D^*_0$. This gap stretches, presumably, up to
$\varepsilon \sim 3\bar\Lambda$.
In general, the situation is very similar to
what occurs with the two-pion contribution in the spectral density induced by
the gluonic current $\alpha_s G^2$ \cite{Shif}, where a strong  low-energy
enhancement is supplemented by an extended gap, and the scale of the
duality violation is quite large.

Note that the $D\pi$ contribution we have calculated  has proper dependence
on parameters
$N_c$ and $N_f$. In the limit $N_c\ra\infty$ it decouples compared to
Eq. (\ref{3}), as it should.  There is no decoupling, however,
if $N_f/N_c$ stays fixed. The pattern we observe presents
another example, similar to Ref. \cite{Shif}, that in the low-energy
domain the $1/N_c$ counting rules should be taken with caution.
There  exist certain mechanisms that can totally upset the $1/N_c$
estimates
at $N_c=3$ (although at academically large $N_c$ the $1/N_c$ counting
will work, of course).

Thus, in the pseudoscalar channel we expect the duality interval to be
$\sim 2\bar\Lambda$, while in the scalar channel
its size is expected to be  larger,  $\sim 3\bar\Lambda$. 
By the duality interval  we mean the following:
the minimal  interval of energy $(0, \varepsilon_0)$ needed
to make the smeared resonance contribution (approximately) equal
to  the quark-gluon one.
In other words, in the smearing 
integrals over the spectral densities from 0 to $\varepsilon_0$ the upper
limit $\varepsilon_0$ should be chosen at $\sim 2\bar\Lambda$ in the 
pseudoscalar channel and  $\sim 3\bar\Lambda$  in the
scalar one.  

We pause here to make a few comments. First, if we consider the sum of the
scalar plus pseudoscalar heavy-light channels (or vector plus axial)
the  quark condensate correction $\langle\bar qq\rangle $ cancels,
leaving us with no hint that the duality
violation scale is larger in this case than in the classical $\rho$ meson sum
rule,  at least at the level of  analyzing the first terms in the operator product 
expansion.
This fact is rather alarming, since in the weak inclusive
decays we deal with the $V-A$ currents, and the above cancellation is quite
typical. The model of Ref. \cite{2} gives no clues in this case
either, since it admittedly omits all effects specifically related to the
spontaneous breaking of the chiral symmetry.
At an observational level we detect here a possible signal
that the scale of the duality violations for $V-A$ currents
may be larger than  can  be inferred from the analysis of
the lowest-dimension condensates. Does this mean that some other
condensates, of higher dimension, must be important? And
how can they be  identified? These questions still remain open. It is clear
that the sum rules for the heavy-light currents have to be reanalyzed
with  emphasis on this aspect.
In particular, an updated analysis
of the scalar (axial) channel is more than in order. One has to replace
the standard ``lowest resonance plus continuum" pattern
by the more complicated picture described above.
The impact seems to be obvious. The  residue of the $D^*_0$ ($D_1$) state
will go down with respect  to the prediction of Ref. \cite{10}. A part of the
spectral density will be pumped out from the resonance into the
non-resonant $D\pi$ background.

The second remark refers to the lattice calculations of the heavy-light
systems (spectra and coupling constants). Pseudoscalar and scalar channels
should be drastically different with respect to inclusion of the
light dynamical quarks. If the pseudoscalar channel seems to be relatively
stable, unquenching quarks and making light quarks really
light must produce a dramatic effect  in the scalar one. Of course, this is not 
the first time
the dynamic light quarks lower the threshold. For instance, in the $\rho$
meson case there is the two-pion cut as well. This state, however,
is only relatively weakly coupled to the current $\bar q\gamma_\mu q$,
since the pions are in the $P$ wave;  therefore, speaking in practical
terms, this contribution is rather unimportant (although at asymptotically
large separations it will dominate anyway). This is not the case for
the current $\bar Q q$, where the $D\pi$ intermediate state
is not only lower in mass than $D^*_0$ ($D_1$), but also strongly coupled to 
the
current.
Therefore, its impact in the two-point function should be essential
and the correlator must drastically change
once the light quarks are unquenched and made light. Studying this problem
on the lattice
seems to be a nice testing ground for various approximations routinely made
within this approach.

\section{Nonleptonic $B$ decays ($b\ra \bar ccs$): outlining the problem}

 Having established the enhancement of the $D\pi$ intermediate state in
${\rm Im}\Pi^{(S)}$ (or $D^*\pi$ in
${\rm Im}\Pi^{(A)}$) near the threshold,  it is natural to  turn to nonleptonic
decays of $B$ mesons, the $b\ra \bar ccs$ transition. Indeed, in this case
the $\bar c \Gamma_\mu s $ current produces  charmed/strange hadronic 
states
with the quantum numbers $1^-$, $0^+$ (the vector current) and
$1^+$, $0^-$ (the axial current). The $1^+$ and $0^+$ channels
were shown above to be responsible for an enhanced production of 
Goldstone mesons in the $S$ wave. Of course, in the actual $B$ decays the 
situation
 is not quite the same as in the limit
$m_Q\ra\infty\, ,\,\,\, m_q\ra 0$ considered in Sect. 2.
Suffice it to mention that the actual values of the $c$ and $s$
quark masses are such that the axial ground-state meson $D_{s1}$,
shown in Fig. 2 at the end of the shoulder (at $\varepsilon
\sim 2\bar\Lambda$), turns out to be almost exactly at its beginning, 
presumably 
barely 
above the threshold of  $D^*K$. The $1/2^+$ charmed strange mesons have  not 
been detected experimentally so far. Moreover, the actual value of $f_B^2$
is also noticeably lower than its static value.  Also,
the energy release in the light particles
in the case at hand is $\le 1.5$ GeV. It is quite clear that the kaon mass
is not negligible in the phase space. For this reason in the 
nonleptonic $B$ 
decays a dedicated analysis is needed. 

First, however, a few remarks regarding  the general situation with
the $b\ra \bar ccs$ transition are in order. This transition came under 
renewed scrutiny recently \cite{5} on purely phenomenological grounds,
in an attempt to find
 a solution
of the ``branching ratio versus $n_c$ problem".  In Ref. \cite{5} it was 
assumed that the theoretical understanding of
$b\rightarrow c\bar u d$ is solid. Then, from
  the measured semileptonic branching ratio of approximately
$10.5\%$, the branching ratio of $b\ra \bar ccs$  was predicted to be close to 
30\%, with the corresponding charm yield $n_c\approx 1.3$. Moreover, using 
the most naive  duality estimates in conjunction with
 the parton model
for $b\rightarrow c\bar cs$, it was
suggested \cite{5} that approximately 1/2 of $b\rightarrow c\bar cs$ 
hadronizes
in the channels of the type $\bar B\rightarrow D^{(*)} \bar D^{(*)}
\bar K X$, and only
the remaining 1/2 goes to $D^{(*)}\bar D^{(*)}_sX$, the channels on which the
attention was focused previously. If so, about 15\% of the $\bar B$ decays
have to end up with the ``wrong" sign $D$'s and $K$'s in the final state.

 This prediction was  confirmed by
 recent 
results from the 
CLEO, ALEPH and DELPHI experiments \cite{7}, reporting the yield 
of the ``wrong" sign $D$'s at the 10\% level. The corresponding value of 
$n_c$ is close to 1.24 \cite{7}. Additionally, ALEPH has recently reported
a value for $n_c$ in $Z\ra b\bar b$ \cite{aleph} of $1.23\pm 0.07$.

Although  the situation can be looked at quite optimistically, serious 
questions are still to be answered. 
The first question is, of course,  purely experimental. As was emphasized
in Ref. \cite{7}, the charm yield $n_c$, as computed in the usual way
from the measurements 
at $\Upsilon (4S)$, remains unchanged:  $n_c = 1.10\pm 0.06$ \cite{17}. 
The contradiction  is obvious, suggesting that the experimental situation is still 
not settled.

We clearly cannot comment on this aspect, and quickly pass to  what
is known theoretically. Since the leading  nonperturbative corrections
are expected to play no significant role in the issue
\cite{6}, the focus of theoretical analysis is shifted towards perturbative 
calculations in $b\ra \bar cc s$. The first dedicated calculation of the
gluon corrections was carried out in Ref. \cite{AP}. The most advanced 
analysis existing today is presented in Refs. \cite{BBSL,volos} (see also 
references 
therein). To quote  representative values of the predicted parameters,
let us assume that $\alpha_s (M_Z) = 0.11$
and $\mu =m_b/2$, in $\overline{MS}$; then \cite{BBSL} $n_c = 1.28$ (the
corresponding BR$_{\rm sl} (B)$ is slightly lower than 11\%). Within a 
somewhat 
different procedure of
treating the ratio of the quark masses $m_c/m_b$ (but the same values of
the parameters as above), the theoretical numbers become  \cite{Pat} $n_c
=1.23$ and BR$_{\rm sl}(B)= 11.5\%$. Thus, it is fair to say that at the current
level of understanding the theoretical prediction for $n_c$ is close to 1.25.
Without the ${\cal O}(\alpha_s )$ correction the charm yield is 1.15.
Thus, the inclusion of the ${\cal O}(\alpha_s )$ gluon correction
enhances BR$(b\ra\bar ccs)$ by a factor $\sim 1.6$. 

In the  approximation of Ref. \cite{BBSL} factorization is explicit.
This statement will be explained in more detail below (Sect. 3), where
we will introduce the factorization hypothesis, one of the key elements of our 
consideration. The impatient reader may turn to Ref. \cite{MBV}
for very clear explanations. Here we just note that, in any perturbative 
calculation respecting factorization,  
the transitions  corresponding to the
vector $\bar c \gamma_{\mu} s$ and the axial $\bar c 
\gamma_{\mu}\gamma_5 s$
currents  necessarily have equal probabilities, provided very small effects
$\propto m_s^2$ are neglected. If so,  the result
 \cite{BBSL} implies that the branching ratios of $b\ra \bar ccs$, with $\bar c 
s$ from the vector and axial currents, 
respectively, are  12 to 13\% each.\footnote {In Ref. \cite{MBV}
some of the ${\cal O}(\alpha_s^2)$ corrections were estimated, and at this 
level 
nonfactorizable terms appear. For the purpose of our analysis one can
safely use factorization, at least as a starting point.}
  
It is clear that the  theoretical predictions \cite{BBSL} discussed above are 
quite compatible with 
the
newest experimental  trend. The basic assumption underlying the theoretical 
analysis
is the validity of the quark-hadron duality. Strong violations of duality in 
$b\ra \bar ccs$  would  destroy the 
predictive power of the existing theory. Although at the moment no sources 
for such large violations were identified \cite{2}, in the absence of a complete 
theory it is obviously desirable to have as many independent confirmations as 
possible.  
We will
analyze  below the transition $b\ra \bar ccs$  from this point of view. 

The transition $b\ra \bar ccs$ is singled out in this aspect for the
following reason. The relative smallness of the energy release,
which alerts one regarding  duality violations, can  be turned into an 
advantage. Indeed, in this case the number of the hadronic channels
saturating the physical decays is not large, and one can try to estimate these
channels one by one to see whether they add up to the quark-gluon result.
As a by-product one could hope to get a more direct  estimate  for the $\bar
B\rightarrow D^{(*)} \bar D^{(*)}\bar KX$ rate.
The goal here is to check whether the 10\% yield reported experimentally
is well understood theoretically,  invoking as few unsubstantiated 
assumptions
as possible.

\section{Nonleptonic $B$ decays ($b\ra \bar ccs$): analysis of exclusive 
modes}

The relevant part of the weak Lagrangian contains two structures,
with the ``direct" and ``twisted" colour flow
\begin{equation}
L=\frac{G_F}{\sqrt{2}} V_{cb}V_{sc}
 \left[ a_1(\bar b\Gamma_\mu c)(\bar c \Gamma_\mu s)+a_2(\bar b
\Gamma_\mu s)
(\bar c\Gamma_\mu c)\right] \, .
\label{7}
\end{equation}
We will disregard the second term, with the twisted color structure, for
the following  reason. First,  the  value of $\vert a_2/a_1\vert $ is
rather small numerically (see e.g. the review paper \cite{17}),
approximately 0.1. Therefore, the square of the second term
contributes at the level of 0.01. The interference with the first
term is suppressed by $N_c$ and is thus  expected to show
up at the level of  corrections $\sim 0.2/3$ in the probability of
the decay modes we are interested in. The estimates to be presented
below have intrinsic theoretical uncertainties of this
order of magnitude or larger.

 Furthermore, in considering the first term it is reasonable to accept,
at least at this stage of the analysis, the factorization
approximation. By this we mean that in treating the hadronic
matrix elements, the $\bar b\Gamma_\mu c $ bracket of
the effective Lagrangian, will be factored out from the
$\bar c \Gamma_\mu s$ bracket. The corresponding hadronic subsystems
are assumed not to communicate with each other through the soft gluon 
exchanges, although inside the subsystems all these exchanges are taken into 
account in full. Note, that we also automatically include all hard gluons (with 
off-shellness larger than
$m_b$), through the factor $a_1$.  The standard theoretical justification for 
the factorization hypothesis is the $1/N_c$ counting. All non-factorizable
contributions are suppressed by $1/N_c$.  Note that the modes with the
hidden charm (e.g. $J/\psi K$) will not concern us here.

Certainly, we are well aware that the non-factorizable
contributions must be present   (see e.g. \cite{Blok});
 their effect is noticeable in the
fine structure of the  nonleptonic decays and in some special modes, but 
otherwise it
is quite modest. For instance, 
in $ B\ra D \bar D_s$
the non-factorizable
contributions were estimated to be less then 10$\%$ of the
factorizable part \cite{bs}. We will ignore the non-factorizable terms in the
present work.

Another theoretical tool which will help us  is the generalized 
small velocity limit, i.e. the assumption that two charmed mesons
we deal with in the $b\ra \bar ccs$ transition, in the final state, are slow.
 Kinematically this means that
\beq
M_B- 2M_D \ll M_D \, .
\label{gsv}
\eeq
This is not the first time the GSV limit is exploited in the context of $B$
decays, see e.g. \cite{MBV,Shif2}. Using the GSV limit,
in combination with factorization, will allow us to disregard
excitations in the $\bar b \Gamma_\mu c$ bracket, and exploit
the well developed formalism of the soft Goldstone mesons
for the modes of the type $D\bar D\bar K$. 

Although in purely theoretical aspect the GSV limit is an excellent
tool, in the actual $B$ decays Eq. (\ref{gsv}) is valid only marginally.
We are neither too close to the GSV point nor too far from it (cf. e.g.
\cite{MBV}). Under the circumstances, 
in  kinematical factors 
we will keep the terms containing the $D$ meson velocities,
while omitting ${\vec v}^2$ terms when they are additionally suppressed. In 
the 
future, with more 
phenomenological input,  this approximation can  be improved, including
all terms quadratic in the (spatial) velocities.

To warm up let us consider two-particle decays. This exercise
is not new (see Ref. \cite{Mannel}), and we repeat it merely
to introduce our notation and explain the choice of  numerical values
of  the relevant parameters.

Thus,  factorization and the GSV limit are starting elements of our analysis.
If so, then the bracket $\bar b\Gamma_\mu c$ is responsible for
the
$\bar B\rightarrow D$ ($B\rightarrow D^*$) transition.
Production of excitations by this bracket -- either resonances  (say, radial
excitations of $D$ and $D^*$),
or nonresonant   states of the type $D\pi$ -- is suppressed by the
velocity  squared of the charmed meson. The $D\pi$ production by 
$\bar b\Gamma_\mu c$ can be studied
within the chiral perturbation theory \cite{USV}. One has to consider pole 
graphs which, in addition to the velocity suppression, are proportional
to the $D^*D\pi$ coupling constant $g$. The latter was calculated within
the QCD sum rules \cite{BBKR}, and turns out to be rather small,
$g\sim 0.3$.  Therefore, here and below we will consistently 
disregard
all contributions to the decay rate that are proportional to $\vec v^2$
and $g^2$.

Limiting ourselves to the $D$ and $D^*$ states in the $\bar b\Gamma_\mu c$ 
bracket
 gives us, by itself,  a predictive power since the form factors
of the $B\ra D^{(*)}$ transitions at zero recoil are normalized
to unity \cite{NW,Shif3}, and near zero recoil are well approximated by
the first derivative of the Isgur-Wise function $\xi$ \cite{IW}.

The second
bracket, $\bar c\Gamma_\mu s$, 
creates sometimes $\bar D_s \,\,{\rm or}\,\, \bar D^*_s$ states, 
sometimes radial excitations of $\bar D_s$ and $\bar D^*_s$,
and
sometimes nonresonant $\bar D^{(*)} \bar K$ pairs. 
The axial current, in its transverse part, can  produce $\bar D_{s1}$ and 
excitations,
while the longitudinal part of the vector current can produce
$\bar D^*_{s0}$ and excitations.
Let us first concentrate on $\bar D_s$ and $\bar D^*_s$.

\vspace{0.2cm}

{\em (i) $B\ra D^{(*)}\bar D^{(*)}_s$. ``Wrong" spin correlations}

\vspace{0.2cm}

The amplitudes for two-body transitions are given by 
\beq
{\cal A}(\bar B\rightarrow D\bar D_s)=\frac{G_F}{\sqrt{2}}
V_{cb}V^*_{cs} \, a_1\,\left( 2\sqrt{M_BM_D}f_{D_s}M_{D_s}\right)
\,\left[\frac{M_B-M_D}{M_{D_s}} \frac{1+vv'}{2}\,\xi (vv')\right]\, ,
\label{A1}
\eeq
\vspace*{0.2cm}

$$
{\cal A}(\bar B\rightarrow D^*\bar D^*_s)=\frac{G_F}{\sqrt{ 2}}
 V_{cb}V^*_{cs}\,
a_1\, \left( 2\sqrt{M_BM_{D^*}}f_{D_s}M_{D_s^*}\right)\times
$$
\beq
\left\{
(\epsilon '\epsilon '')\frac{1+vv'}{2} +\left[
-\frac{1}{2}(\epsilon ' v)(\epsilon '' v') +
\frac{i}{2}\epsilon_{\mu\nu\rho\lambda} \epsilon '_\mu\epsilon ''_\nu 
v_\rho 
v'_\lambda\right]
\right\}\xi (vv')\, ,
\label{A2}
\eeq
\vspace*{0.2cm}

$$
{\cal A}(\bar B\rightarrow D\bar D_s^*)=\frac{G_F}{\sqrt{2}}
V_{cb}V^*_{cs} \, a_1\times
$$
\beq
\left(2 \sqrt{M_BM_D}f_{D_s^*} M_{D_s}\right)
\,\left[ \frac{M_B+M_D}{2M_{D}}\, (\epsilon '' v)\, \xi (vv') \right]\, .
\label{A4}
\eeq
\vspace*{0.2cm}

$$
{\cal A}(\bar B\rightarrow D^*\bar D_s)=\frac{G_F}{\sqrt{2}}
V_{cb}V^*_{cs} \, a_1\times
$$
\beq
\left( 2\sqrt{M_BM_{D^*}}f_{D_s^*} M_{D_s^*}\right)
\,\left[ \frac{M_B+M_{D^*}}{2M_{D_s}}\, (\epsilon ' v)\, \xi (vv') \right]\, .
\label{A3}
\eeq
\vspace*{0.2cm}\\
Here we omitted inessential overall phase factors in front of some amplitudes.
The constants $f_{D_s}$ are defined as , 
$$
\langle \bar D_s |\bar c\gamma_\mu\gamma_5 s|0\rangle
= -i f_{D_s}M_{D_s}v''_\mu\, , \,\,\,
\langle \bar D_s^* |\bar c\gamma_\mu s|0\rangle =
f_{D_s}M_{D_s^*}\epsilon''_\mu\, .
$$
The  pseudoscalar and 
vector constants  are identical in the limit $m_c\ra\infty$, but 
may differ a little, due to preasymptotic terms, for the actual $c$ quarks.
For simplicity the difference between the pseudoscalar and vector constants 
are neglected
within any hyperfine multiplet. Anyway, we do not know them 
to that accuracy. 
Moreover,
$v$ is the four-velocity of the decaying $B$ meson, $v'$ is the four-velocity
of the $D^{(*)}$ meson and $v''$ is the four-velocity of $\bar D_s^{(*)}$,
$$
vv'=\frac{M_B^2+M_D^2-M_{D_s}^2}{2M_BM_D}\;,\;\;\;
vv''= \frac{M_B^2+M_{D_s}^2-M_D^2}{2M_BM_{D_s}}\;,\;\;\;
v'v''=\frac{M_B^2-M_D^2-M_{D_s}^2}{2M_D M_{D_s}}\, ,
$$
and likewise for other decays,
$\epsilon '$ and $\epsilon ''$ are the polarization vectors of $D^{*}$
and  $\bar D_s^{*}$, respectively.

In the limit of slow $D$'s, when $vv',\,vv'',\;v'v'' \ra 1$,  
these expressions essentially 
simplify. Thus, in Eq. (\ref{A1}) the square bracket becomes unity, and
in Eqs. (\ref{A2}),  (\ref{A4}) and (\ref{A3})  the square brackets tend to 
zero --  as $\sqrt{vv''-1}$ in Eq. (\ref{A4}) and as $\sqrt{vv'-1}$ in 
Eq. (\ref{A3}). 
Thus, in this limit we have a rigid spin 
correlation.
If $D$
is produced by the $\bar bc$ bracket,  the $\bar c s$ bracket  will
produce   a pseudoscalar $\bar D_s$;
if $D^*$ is produced by the $\bar bc$ bracket we get, in association,  a vector
$\bar D^*_s$.
This is because in this limit the $B\ra D$ transition is caused by the zeroth 
(time) component of the current, whereas $B\ra D^*$ is due to the spatial 
component.
The ``wrong"-spin transitions, Eqs. (\ref{A4}) and (\ref{A3}), switch off 
in this limit.
Table 1 gives the values of the parameters $y=vv'$, $v'v''$ and $vv''$ 
for various transitions in the
ground state and excited resonances. In the GSV limit all these kinematic 
parameters reduce to unity. 
\begin{table}
\begin{center}
\caption{ }
\vspace{0.1in}
\begin{tabular}
{|c|c|c|c|c|}\hline
Decay &$y$&$v'v{''}$&$vv{''}$&${\rm block}\,\,{\rm factor}$\\ \hline
$B\rightarrow D\bar D_s$&1.39 &2.79&1.36&1.50\\ \hline
$B\rightarrow D\bar D^*_s$&1.36&2.53&1.29&1.25\\ \hline
$B\rightarrow D^*\bar D_s$&1.32&2.52&1.33&.93\\ \hline
$B\rightarrow D^*\bar D^*_s$&1.29&2.28&1.27&$\sqrt{3}\cdot 0.94$\\ \hline
$B\rightarrow D\bar D_s'$  &1.25 & 1.82 & 1.13&1.22\\ \hline
$B\rightarrow D\bar D_s^{*\prime} $ & 1.21 & 1.64& 1.10& 0.79\\ \hline
$B\rightarrow D^*\bar D'_s$ & 1.19 &1.64  &1.11&0.58 \\ \hline
$B\rightarrow D^*\bar D^{*\prime}_s$
&1.15&1.47&1.08&$\sqrt{3}\cdot 1.04$\\ \hline
$B\rightarrow D\bar D_s''$ & 1.10 &1.28 &1.04 &1.07 \\ \hline
$B\rightarrow D\bar D_s^{*\prime}$ & 1.05 & 1.14& 1.02& 0.37\\ \hline
$B\rightarrow D^*\bar D_s''$ & 1.05 &1.14  &1.02&0.28 \\ \hline
$B\rightarrow D^*\bar D_s^{*\prime}$ & 1.006 &1.02 &1.002 &$\sqrt{3}\cdot 
1.09$\\ 
\hline
\end{tabular}
\end{center}
\end{table}

This table also  gives the values of ``block factors". 
They are defined as 
$$
\mbox{block factor} = \left( \frac{{\cal A}^2}{|\frac{G_F}{\sqrt{2}}
V_{cb}V^*_{cs} \, a_1\,
\left(2 \sqrt{M_BM_D}f_{D_s}M_{D_s}\right)
|^2}\right)^{1/2}
$$
where the summation over the polarizations of $D_{(s)}^*$
is implied, where applicable; for the excited $D_s$ states considered below 
$M_{D_s}$ and
$f_{D_s}$ in the
denominator will be understood as the mass and the coupling of the 
respective pseudoscalar.
We included here the IW function $\xi(vv')$, see below.
In the first decay the block factor is just the 
square bracket in Eq. (\ref{A1}).

We see that in the $\bar B \ra D\bar D_s$ transition
the block factor which is equal to unity in the GSV limit, is actually close
to $1.5$, while in the ``wrong" spin correlation transition $\bar B \ra D\bar 
D_s^*$
the block factor, which vanishes in the GSV limit is actually close
 to $1.25$.  This is a manifestation of the fact that in the two-body 
decays
into the lowest-lying $D$'s the GSV approximation is not too good,
as was expected, of course. 

 Table 1 shows  that the GSV approximation becomes much better for excited 
$\bar D$'s where the wrong spin transitions are 
indeed suppressed. The GSV approximation, quite naturally, 
significantly improves in
the three-particle decays as well, where a part of the overall energy release
goes to create, additionally,  the kaon mass, and the remainder is shared by 
three particles, not two. 

If instead of $\bar D_s$ and $\bar D_s^*$ we have their radial excitations,
Eqs. (\ref{A1})--(\ref{A3}) are modified in a minimal way.
Apart from the masses and the kinematical factors, 
one must replace $f_{D_s}$ by the corresponding coupling.

For numerical estimates of the decay rates
we need to fix various parameters. First, we take
$f_{D_s}\approx 200 \MeV$ \cite{12,BE}. Now,  the
Isgur-Wise function $\xi$ must be evaluated  at the proper 
recoil values,  which are given in Table 1. 
We use the linear approximation for the Isgur-Wise function 
\beq
\xi (vv') = 1-\rho^2 (vv'-1) \, ,
\label{iwlin}
\eeq
where $\rho^2$ is the slope parameter. Its numerical value is more or less
known, and we put
$$
\rho^2 = 0.7\, .
$$
This value is
compatible with the experimental data \cite{BR} and with the QCD sum rule
calculations \cite{20}.  Finally, $a_1\approx 1.1$. 

The relevant decay rates are 
\beq
\Gamma (B\rightarrow D\bar D_s)\;=\;\Gamma_0\, \left(
\frac{M_B-M_D}{M_{D_s}}\, \frac{1+vv'}{2} \right)^2
\, \xi^2 \, ,
\label{G1}
\eeq
\vspace*{0.2cm}

\begin{equation}
\Gamma (B\rightarrow DD^*_s)
\;=\; 
\Gamma_0^*\,
\frac{(M_B+M_D)^2}{4M_D^2}\, ((vv'')^2-1) \, \xi^2 \, ,
\label{cf}
\end{equation}
\vspace*{0.2cm}

\begin{equation}
\Gamma (B\rightarrow D^*\bar D_s)\;=\;\Gamma '_0\,
\frac{(M_B+M_{D^*})^2}{4M_{D_s}^2}\, ((vv')^2-1)\,  \xi^2\, ,
\label{df}
\end{equation}
\vspace*{0.2cm}

\beq
\Gamma (B\rightarrow D^*\bar D^*_s)=3\Gamma_{0}'^*\, 
\left[1+\frac{vv'-1}{3} +\frac{(vv')^2-1}{4} - \frac{(v'v'')^2-1}{12}
+\frac{(vv')(vv'')-1}{3}
\right]\,\xi^2 .
\label{G2}
\eeq
Here 
$$
\Gamma_0 = \frac{G^2_F}{4\pi}
\vert V_{cb}V_{cs}\vert^2a_1^2
\, f^2_{D_s}\, M_DM_{D_s}^2 \frac{|\vec p |}{M_B}\, ,
$$
and $\Gamma_0*$, $\Gamma_0'$, $\Gamma_0'^*$
are the same with the $D$ ($D_s$) replaced by $D^*$ ($D_s^*$) for asterisk, and  
$\vec p$ is the c.o.m.  momentum of the produced mesons.  
Numerically the relevant branching ratios are given in  Table 2.
\begin{table}
\begin{center}
\caption{ }
\vspace{0.1in}
\begin{tabular}
{|c|c|}\hline
Decay &${\rm Branching}\,,\;\%$\\ \hline
$B\rightarrow D\bar D_s$&1.1\\ \hline
$B\rightarrow D\bar D^*_s$&0.87\\ \hline
$B\rightarrow D^*\bar D_s$&0.45\\ \hline
$B\rightarrow D^*\bar D^*_s$&1.5\\ \hline
${\rm Total}\;\; {\rm Br}\,, \;\%$ &  $4.0$ \\ \hline
\end{tabular}
\end{center}
\end{table}

So far we basically
repeated the calculations of Mannel {\em et al.} \cite{Mannel}; our estimate of 
the Isgur-Wise function
is different, and our value of $f_{D_s}$ is essentially lower than that
accepted in Ref.  \cite{Mannel}.  
We note that the total sum of the above four modes 
amounts  to $\sim 4\%$   in the branching
ratio. None of these modes produces ``wrong" sign $D$'s. 
The  above numerical  results   are in agreement with
the experimental data \cite{21},
\begin{equation}
\Gamma (B^0\rightarrow
D\bar D_s)= 0.7\pm 0.4\%\; ,\; \;\;\;
\Gamma (B^0\rightarrow D^*\bar D^*_s) =
1.9\pm 1.2\%\; ,
\label{11a}
\end{equation}
\begin{equation}
{\rm Br} (B^0\rightarrow D^*\bar D_s)=1.2\pm 0.6\%\; ,\;
{\rm Br} (B^0\rightarrow D\bar D^*_s)=2\pm 1.5\%\, ,
\end{equation}
\begin{equation}
{\rm Br}
(B^+\rightarrow D\bar D_s)=1.7\pm 0.6\%\; , \;\;\;\;
{\rm Br} (B^+\rightarrow D^*\bar D^*_s) = 2.3\pm 1.4\%\; ,
\label{11l}
\end{equation}
\begin{equation}
{\rm Br} (B^0\rightarrow D^*\bar D_s)=1\pm 0.7\%\;
 , \; {\rm Br}(B^0\rightarrow D\bar D^*_s)
=1.2\pm 1\%)\, .
\end{equation}

Encouraged by this success we can proceed to other exclusive modes
which have  not been discussed in the literature so far.
\vspace*{0.2cm}

{\em (ii) Modes with the radial excitations of $\bar D_s^{(*)}$}
\vspace*{0.2cm}

We got used to the fact that the radial excitations are coupled
to  corresponding currents significantly weaker than
the ground-state mesons with the given quantum numbers.
This is, for instance, the case with the $J/\psi$ and $\Upsilon$ mesons
whose coupling to the current $\bar Q\gamma_\mu Q$ decreases
with the excitation number.
In the decays $ B\ra D^{(*)} \bar D_s^{(*)'}$ (where the prime denotes
the first radial excitation and the double prime will denote the second 
excitation)
we expect the opposite trend. This is a specific feature of the
$\bar Q q$ current. The corresponding parton spectral density grows
quadratically with the energy $\varepsilon$, see Sect. 2. If duality
takes place, the residues should be roughly proportional to the corresponding
areas under the parabola in Fig.~1. To get an idea of the excited state 
contribution
we will assume that
$M_{D_s'} \approx 2.6 \GeV$, $M_{D_s''} \approx 3.1 \GeV$,
$M_{D_s^{{*\prime}}} \approx 2.75 \GeV$ and  $M_{D_s^{*\prime\prime}} 
\approx 3.25 \GeV$, 
 i.e. are equidistant.  Imposing the duality condition in the form 
$$
\int_{M_{k-1}}^{M_k}\; dM\;\Im \Pi_{\rm pert}(M)\;=\; \int\;
dM\;\Im \Pi_k(M)\;,\;\;\; \Im \Pi_k(M)\;\propto \; M_k f_k^2 \,\delta(M-
M_k)
$$
we then obtain
\beq
\frac{f_{D_s'}^2 M_{D_s'}}{f_{D_s}^2 M_{D_s}} 
\approx \frac{27-8}{8} \approx 2.5\;,\;\;\;\;
\frac{f_{D_s''}^2 M_{D_s''}}{f_{D_s}^2 M_{D_s}}
\approx \frac{64-27}{8} \approx 4.5\;.
\label{dual}
\eeq
Note that we use here an ``extended" definition of the excited resonances,
which need not coincide with the standard Breit-Wigner peaks,
with the backgrounds subtracted. Rather, we include backgrounds,
collecting, say, in $D_s'$ everything with the invariant mass lower
than $\approx 2.3 \GeV$ (except the ground state $D_s$, of course). This is a 
natural
operational definition from the point of view of the QCD practitioner.

The estimates in Eq. (\ref{dual}) are crude and intended
only for the purpose of orientation. However, it is a model for the
spectral density that exactly respects duality to the given approximation, 
and, thus,
allows simultaneous estimates of  its violations  in the actual decay
width,  where only  a limited energy release is available. The contribution of 
the
last open multiplet is actually a reasonable  estimate of the scale of  duality 
violations that might occur. 
On the practical side, to allow a measure of theoretical 
uncertainty in the excited state residues, we will allow  the ratios to float,
\beq
(f_{D_s'}^2 M_{D_s'})/(f_{D_s}^2 M_{D_s})\; =\; 2.5 \, x \; , \;
\;\;\;  (f_{D_s''}^2 M_{D_s''})/(f_{D_s}^2M_{D_s}) \;= \;4.5 \, x
\eeq
where $x$ will be varied between 0.6 and 1. The values of $x<1$ reflect a
relative enhancement of the lower end of the spectral density by the 
perturbative and, to some extent, condensate effects. Also the higher end
of the spectral density is somewhat suppressed since at these energies 
relativistic effects become important, and
temper the $\varepsilon^2$ growth of the spectral density
characteristic to the non-relativistic limit.
As we will see later on, the data are more consistent with $x\sim 0.6$.

In Table 3 the branching ratios for the two-particle  decays into excited states
are quoted for three different values of $x$ .
\begin{table}
\begin{center}
\caption{}
\vspace{0.1in}
\begin{tabular}
{|c|c|c|c|}\hline
Decay & ${\rm Br}, \%\,\,  (x=0.6)$ & ${\rm Br}, \%\,\,
 (x=0.8)$ & ${\rm Br}, \%\,\, (x=1)$\\ \hline
$B\rightarrow D\bar D_s'$  & $1.1$ & $1.5$ & $1.9$\\ \hline
$B\rightarrow D\bar D_s^{*\prime}$ & $0.5$ & $0.65$ & $0.8$\\ \hline
$B\rightarrow D^*\bar D_s'$ & $0.25$ & $0.35$ & $0.4$\\ \hline
$B\rightarrow D^*\bar D_s^{*\prime}$ & $2.4$ & $3.2$ & $4.0$\\ \hline
${\rm Total}\;\;\bar D_s^{(*)\prime}\; 
{\rm Br}, \%$ & $4.3$ & $5.7$ & $7.1$ \\ \hline
$B\rightarrow D\bar D_s''$ &$1.2$ &$1.6$ & $2.0$ \\ \hline
$B\rightarrow D\bar D_s^{*\prime\prime}$ &$0.1$ &$0.15$ & $0.2$\\ \hline
$B\rightarrow D^*\bar D_s''$ &$0.06$ &$0.08$ & $0.1$\\ \hline
$B\rightarrow D^*\bar D_s^{*\prime\prime}$ &$1.1$  &$1.5$  & $1.9$\\ \hline
${\rm Total}\;\;\bar D^{(*)\prime\prime}\; 
{\rm Br}, \%$ & $2.5$ &$3.3$ & $4.1$ \\ \hline
${\rm Total}\;\;\bar D^{(*)\rm excit}\;{\rm Br},\%$&$6.8$&$9.0$&$ 11.3$\\ 
\hline
\end{tabular}
\end{center}
\end{table}
Altogether we get that 
$$
{\rm Br} (B\rightarrow D^{(*)} + \mbox{excited}\,\,  D_s^{(*)}) \sim 7\;
{\rm to }\, 12\%\; .
$$
Let us discuss the decay pattern
of these excitations. This issue is interesting 
not only in connection with the ``wrong" sign $D$'s in the $B$ decays,
but by itself as well. 

$\bar D_s'$ can decay neither into $\bar D\bar K$ nor into $\bar D^*\bar K$. 
The second mode is presumably below the threshold. Even if it is slightly 
above
the threshold, the small energy release, in conjunction with the $P$ wave
nature of the decay, strongly suppresses it. Thus, the dominant mode
is expected to be $\bar D_s \pi\pi $. Therefore, the $\bar D_s'$
production does not generate the ``wrong" sign $D$'s.

On the other hand, $\bar D_s''$ will presumably decay predominantly
into 
$\bar D^{*}\bar K$. One third of the above goes into   $\bar D_s^{*} \eta$. 
Some $\bar D\bar K^{*}$ are also possible. The latter mode  also leads to
a ``wrong" sign $D$. Its presence may somewhat distort, however, our 
estimate of $\bar D_s^{*} \eta$. 
A competition 
from the $\bar D_s 
\pi\pi $
mode is generally  possible, but unlikely to be essential: the $\bar 
D_s'' \ra \bar D_s 
\pi\pi $ is a three-body decay, with
the additional suppression due to the pion momenta.  (It can proceed via 
an $S$-wave
$\pi\pi$ resonance; however, the $\sigma$ meson is too broad and leads only 
to a
moderate enhancement, whereas higher resonances do not have much of a 
phase 
space.)

As for $\bar D_s^{{*\prime}}$ and $\bar D_s^{{*\prime\prime}}$, their 
dominant decay  mode is likely to 
be  $\bar D^{(\prime)}\bar K$ and $\bar D_s^{(\prime)} \eta$. 
Thus, in the three latter cases we end up with
the ``wrong" sign $D$'s. Due to $SU(3)_{\rm fl}$ symmetry the yield of
$\eta$ is three times smaller than that of $\bar K$'s. The total branching of 
the 
corresponding modes is $5$  to $ 9\%$. The  $D\bar D_s\eta$ yield from this 
mechanism is $1.2$ to $2.3\%$.    

\vspace{0.2cm}

{\em (iii) Nonresonant $D^{(*)}\bar D^{(*)} \bar K$ modes}

\vspace{0.2cm}

So far we discussed the $1^-$ and $0^-$ channels of the
$\bar c s$ current. 
Now let us proceed to the positive-parity channels.

As is clear from Sect. 2 the nonresonant $S$ wave kaons
are produced by
$\bar c\gamma_\mu\gamma_5 s$ in association with $\bar D^{(*)}$.  The 
longitudinal part
of the current produces $\bar D \bar K$, while the
transverse component produces $\bar D^* \bar K$.
We will assume that the kaons can be treated as soft
Goldstone mesons. Needless to say,  factorization is implied too.

Then, we have to write out the currents $\bar b\Gamma_\mu c$
and $\bar c\Gamma_\mu s$ in the chiral/heavy quark theory.
The second current is given  in Ref. \cite{16}.  The amplitude
of the kaon ``bremsstrahlung" in the transition
$B\ra D^{(*)}\bar D^{(*)}\bar K$ consists of two parts --  contact  and 
pole (see e.g. graphs of Figs.~1 and 2, respectively,  in Ref. \cite{16}; see
also \cite{USV}). It is not 
difficult to check that all pole graphs vanish in the GSV limit, when $D$'s 
spatial
velocities are set equal to zero. Additionally they are 
numerically suppressed by $g$, the $D^*D\pi$ coupling constant,
which, according to Ref. \cite{BBKR}, is about 0.3. Therefore, at the current 
level 
of accuracy, we prefer to omit the pole graphs altogether. In the future, as 
confidence in the numerical value of $g$  grows and estimates of the excited 
meson couplings  acquire  better accuracy, it will be necessary to return to 
the issue and include the pole graphs in the theoretical predictions.

With the pole graphs discarded, calculating the decay rates of the
above transitions becomes trivial. Indeed, the ratio of the
amplitude $D^{(*)}\bar D^{(*)} \bar K$ in the limit $p_K\ra 0$ to the 
corresponding 
two-body amplitude $D^{(*)}\bar D^{(*)}_s $ is given by $i/f_\pi$.
(Note that we will consistently neglect all SU(3)$_{\rm fl}$ breaking effects
everywhere, except the phase spaces.
Accounting for SU(3)$_{\rm fl}$ breaking effects in the amplitudes in the first 
order in $m_s$ is possible;
we will defer this exercise as well.) Since the kaons are produced in the $S$ 
wave, and the contact amplitudes depend neither on the kaon energy
nor on the angles, the ratio of probabilities is given merely by
the ratio of the three-body to two-body phase spaces $V_{2,3}$,
\beq
\frac{\Gamma (\bar B\ra D^{(*)}\bar D^{(*)} \bar K)}{\Gamma
(D^{(*)}\bar D^{(*)}_s)} = \frac{1}{f_\pi^2}\, \frac{V_3}{V_2} \, .
\label{ddkrat}
\eeq
The three-particle phase space is conveniently written out e.g. in Ref.
\cite{22}.

Note that the above estimate is valid only for the ``right" spin correlation 
modes, $D\bar D\bar K$ and $D^{*}\bar D^{*} \bar K$. The ``wrong" spin 
correlation modes vanish in the GSV limit, for the same reason as was 
explained above, and cannot be computed without inclusion of the pole
graphs. Their role, however, is significantly reduced
in the three-particle decays compared to the two-particle case,
since in the three-particle decays we are closer to the GSV limit.

It is convenient to present the ratio $V_3/V_2$ as follows:
\beq
\frac{V_3}{V_2} = \frac{M_B^2}{32\pi^2}  \, \frac{I_3}{I_2}\, ,
\label{vv}
\eeq
where the first factor corresponds to the ratio of the phase
spaces with all {\em massless} final particles. The dimensionless
factors $I_3$ and $I_2$ reflect the finite masses: 
$$V_3\,=\,M^2_BI_3/(256\pi^3)\;,\;\;\;
V_2\,=\,\frac{1}{4\pi}\vec p/M_B=\frac{I_2}{8\pi}\;.
$$
Numerically $I_3$ are summarized in Table 4, while the branching ratios 
for three-particle yield for different nonresonant decay modes are given in 
Table 5. 
\begin{table}
\begin{center}
\caption{}
\vspace{0.1in}
\begin{tabular}
{|c|c|}\hline
Decay &$I_3$\\ \hline
$B\rightarrow D\bar D\bar K\;(\eta)$ &$0.073\;(0.068)$\\ \hline
$B\rightarrow D^*\bar D^*\bar K\;(\eta)$& $0.039\;(0.035)$\\ \hline
$B\rightarrow D\bar D'\bar K\;(\eta)$&$0.014\;(0.002)$\\ \hline
$B\rightarrow D^*\bar D^{*\prime}\bar K\;(\eta)$&$0.002\;(0.001)$\\ \hline
\end{tabular}
\end{center}
\end{table}

Using the numerical values for $I_3$ 
one can easily estimate the corresponding 
branching ratios:
\beq
\frac{\Gamma (\bar B\ra D\bar D \bar K)}{\Gamma
(D\bar D_s)} \approx 0.5 \, , \,\,\, 
 \frac{\Gamma (\bar B\ra D^{*}\bar D^{*} \bar K)}{\Gamma
(D^{*}\bar D^{*}_s)} \approx  0.7 \, ,
\eeq
and 
\beq
\frac{\Gamma (\bar B\ra D\bar D' \bar K)}{\Gamma
(D\bar D_s')} \approx 0.2\, , \,\,\, 
 \frac{\Gamma (\bar B\ra D^{*}\bar D^{{*\prime}} \bar K)}{\Gamma
(D^{*}\bar D^{{*\prime}}_s)} \approx 0.04\, .
\eeq
Here we accounted for the fact that one can have  two $\bar K$'s, of different 
charge, in each process at hand, and incorporated the block factors for the
two-body modes. The relevant branching ratios, computed directly, are 
collected in Table 5. 
\begin{table}
\begin{center}
\caption{}
\vspace{0.1in}
\begin{tabular}
{|c|c|}\hline
Decay & ${\rm Br}\,, \%\;\, (x=1)$\\ \hline
$B\rightarrow D\bar D\bar K$ & $0.5$\\ \hline
$B\rightarrow D^*\bar D^{*}\bar K$ & $1$\\ \hline
$B\rightarrow D\bar D'\bar K$ & $0.3$\\ \hline
$B\rightarrow D^*\bar D^{{*\prime}}\bar K$ & $0.15$\\ \hline
\end{tabular}
\end{center}
\end{table}

 Altogether we see that the total branching for four possible 
nonresonant channels is $\sim 2\%$. In the same approximation it is easy to 
estimate
the yield of nonresonant $\bar D^{(*)}_s\eta$ instead of $\bar D^{(*)}\bar K$.
In the SU(3)$_{\rm fl}$ limit it is 1/3 of the kaon yield.
Taking account of the differences in the phase spaces we get
Br($ B\rightarrow D\bar D_s\eta ) \sim 0.3$ to $0.5\%$.

Altogether the nonresonant channels give $\sim 2-3\%$ of the total 
branching  ratio.

\vspace{0.2cm}

{\em (iv) Other modes with nonresonant Goldstone mesons}

\vspace{0.2cm}

Invoking the chiral/heavy quark technique and factorization in the same vein 
as above, it is possible to calculate  amplitudes with two soft Goldstone mesons
(nonresonant), e.g.
$D \bar D \bar K\pi$
or  $D\bar D_s\pi\pi$. For instance, the  $\bar D^* \bar K\pi$ state will be 
produced by the
vector $\bar c\gamma_\mu  s$ current, and all relevant matrix elements are 
already known
in the literature \cite{B}. A number  of  pole graphs are to be taken into 
account. We did not attempt this calculation, because
the four-particle phase space in the processes at hand is prohibitively small.
The decays with two nonresonant Goldstone mesons are rare.

\vspace{0.2cm}

{\em (v) Resonances with the quantum numbers $0^+$ and $1^+$ in
the $\bar c s$ channel }

\vspace{0.2cm}

Since the non-resonant contributions in the above channels
turn out to be rather small, well below the 12\% level following from
the quark-gluon calculation (plus duality), it is natural to conclude
that $D_{s1}$ and $D_{s1}'$ (and $D_{s0}^*$ and $D_{s0}^{{*\prime}}$)
play a significant role. The mass of $D_{s1}$ is expected to be $\sim 2.5$ GeV, 
and it
is natural to expect that the mass of $D_{s1}'$ is close to 3 GeV.
The masses of $D_{s0}^*$ and $D_{s0}^{{*\prime}}$ are  close to those above.

The estimates of the corresponding yield can be approached 
in a way similar to
what has been done for the vector (pseudoscalar) channel. (The effect of the 
$s$-quark mass may be harder to take into account
properly, since, as we have seen,  the
significant chiral kaon contribution gets suppressed, which may affect the
residues of the resonances.) We  use the following simplified estimate.
The rate of $D_{s1}$ 
is expected to be approximately the same as that of $D_s^{*\prime}$. The first
excitation of $D_{s1}$ is produced roughly with the same rate as the
second excitation of $D_s^*$. The same is valid with respect to
$D_{s0}^*$ and $D_{s0}^{{*\prime}}$ relative to the excitations of $D_s$.

One may expect that  
all these resonances decay in the $\bar D^{(*)}\bar K$  mode; in some cases
 $K^* $ may appear, instead of $K$. This
will still lead to the ``wrong" sign $D$.
Then one gets $\sim 7$ to $11\%$  depending on the  value of $x$.

\vspace{0.2cm}
{\em (vi) Excited D states produced by $\bar b\Gamma_\mu c$ bracket}
\vspace{0.2cm}

Above we considered only decays of $B$ into $D^{(*)}$ and some radially 
excited $\bar D_s$
states. There may  also be decays into the excited states of $D$. The
total yield for such decays with the radial excitations is quite small. 
There exist two possibilities: one can produce from the $\bar b\Gamma_\mu 
c$ 
current either radial excitations of $D^{(*)}$, or $P$-wave mesons.

In the first case
the corresponding 
partial widths are proportional to the square of the Isgur-Wise function 
of the transition between $B$ and the radial excitations of $D$. However, the 
corresponding IW function vanishes as $vv'-1$. 
Thus the corresponding partial widths are suppressed by a factor
$(vv'-1)^2\sim \vec v^{\prime \,4}$
relative to the transitions into $D^{(*)}$. A
direct calculation shows that if one takes the masses of the excited $D$ states 
to lie uniformly below the  masses of the excited $\bar D_s$ states by 150 
MeV,
the corresponding recoil values of $vv'-1$ spread from 0 to 0.2. Thus the 
corresponding partial widths are suppressed by a factor of order 0.04
relative to the decays into the non-excited states. Our estimate would 
show that the corresponding contribution is less than $1\%$; although 
a larger effect is expected from the power corrections \cite{USV,optical},
up to 10 to $20\%$ of the elastic transition, they still seem to be small.

Higher rates can be due to  the $P$-wave mesons from the $\bar 
b\Gamma_\mu c$ current; in this case  the rates are 
only linear in $vv'-1$. An upper bound (and a reasonable estimate) of them 
can
be obtained by merely calculating the sum over all  decay modes with 
$D$ and
$D^*$, with a ``faked" $\rho=0.25$ (or with the 
excitation-free IW function $\xi_0(vv')=((1+vv')/2vv')^{1/2}$),  through
the sum rules for 
the transition amplitudes \cite{optical}. We then  compare the result obtained
to that with the ``actual'' $\xi$ of \eq{iwlin}:
$$
{\rm Br}(B\ra D^{\rm excit}+X_{\bar{c}s})\; \approx \;
{\rm Br}(B\ra D^{(*)}+X_{\bar{c}s})\vert_{\xi=\xi_0}-
{\rm Br}(B\ra D^{(*)}+X_{\bar{c}s})\vert_{\xi=1-\rho^2(vv'-1)}\;\;.
$$
Since for higher excitations of $D_s^{(*)}$ the phase space is limited, this
relation only  overestimates somewhat the associated  production of excited
$D^{(*)}$.
In this way we arrive at the conclusion that the contribution of the excited 
$D^{(*)}$ states from the $\bar b\Gamma_\mu c$ bracket is 
$\lsim 3\%$.

\section{Conclusions}

The existence of the Goldstone mesons results
in a very peculiar pattern of the heavy-light spectral densities,
which turn out to be drastically different, say, in the vector and axial
channels. By treating kaons as soft,  we are able to calculate the nonresonant
$D^{(*)}\bar D^{(*)}\bar K$ ($D^{(*)}\bar D_s^{(*)} \eta$) yield  in $B$
decays.

Altogether, summing all two-body and three-body decays modes, we get 
$23$ to $31$\% for the total $b\rightarrow \bar ccs$ yield. The first number  
corresponds to $x=0.6$ and the second   to $x=1$.
Moreover, if our  assumptions about the decay modes of $D_s''$, 
$D_s^{*\prime}$,  $D_s^{*\prime\prime}$, $D_{s1}$, $D_{s1}'$,
$D_{s0}^*$ and $D_{s0}^{*\prime}$ are justified, they lead predominantly to 
the
``wrong" sign $D$'s. The yield of the ``wrong" sign $D$'s then comes out
to be 
13 to $ 19\%$.\footnote{We assume here, rather arbitrarily, that the excited
$D^{(*)}$ mesons from the $\bar b c$ bracket are produced at the level of 2\%,
and this yield is split equally between the ``wrong" sign $D$'s and 
$D_s$.} The yield of the ``right sign" $D$'s is estimated to be 10 to $ 12\%$;
about 4\% comes from the modes $D^{(*)}\bar D_s^{(*)}$, and about the same 
from 
the 
modes with $\eta$ in the final state.
Thus, not only we confirm, exploiting  a different line of reasoning, the
conclusion of Ref. \cite{5} regarding the abundance of the ``wrong" sign
$D$'s, but we encounter a menace of ``overshooting": the yield of
the ``wrong" sign
$D$'s naturally tends to be too high.

Under the circumstances it is tempting to calibrate theoretical predictions
by using 
the  CLEO and ALEPH data on  the ``wrong" sign
$D$'s, 
${\rm Br}\,(B\rightarrow D^{(*)}\bar D^{(*)}\bar K )=10\pm 4\%$. 
Combining this result with our estimates we see that the value of
$x\approx 0.6$ is somewhat 
preferred.  If so, the theoretically  expected branching ratio
of the ``wrong" sign
$D$'s is about 13\%, that of the ``right sign"
$D$'s is about 10\%, and the total branching ratio of the
$\bar c c s$ channel is about 23\%.
Thus, we see that 
our model of 
saturation of inclusive data with two- and three-particle exclusive modes 
is perfectly consistent with the duality-based  prediction 
of Ref. \cite{BBSL}, $\sim 23\%$
or slightly higher. It is quite remarkable that the  exclusive estimates 
presented above do not contradict
the quark-hadron duality. This is not too surprising, of course, since in
modelling the resonance yields we respected the constraints imposed by
QCD plus factorization. The accuracy of this conclusion is not high, though, 
at the level of $30\%$. 

 As an experimental confirmation of our approach the $D\bar D_s\eta +...$ 
yield will have to be seen at  the $4\%$ level.

The predictions for the absolute rates made above  do not pretend to have too 
high an accuracy; suffice it to mention that the rates are proportional to 
$f^2_{D_{(s)}}$, and this quantity 
carries a noticeable uncertainty. However, it seems certain that,
 contrary to naive expectations, the modes with $\bar D_s$ (the ``right" sign 
$D$'s) 
constitute a smaller fraction of  the $b\ra c\bar c s$ channel. This seems
to be a rather general feature not depending on  details of the approximations 
we rely on. The spectral density emphasizes  larger invariant masses of the
$\bar c s$ system, which then predominantly generates decay chains with 
strangeness and 
charm
separated.

In our calculations we employed factorization,  which must be violated at 
some 
level. However, if the cross-talking between two
currents $\bar c s$ and $b\ra c$ appears to be strong, then one must expect
dissolving  $D_s$ into separate charm and strange particles from the very
beginning.

It
is worth noting that the relative yield of $D_s^{(*)}$ is more sensitive to the 
strange quark mass,
than  the overall $\bar c c s$ yield.

We point out that, although our calculations do not seemingly suggest a
prominent 
mechanism for differentiating the hadronic widths of $B$ and $\Lambda_b$ 
in the $b\ra c\bar c s$ channel, where the calculations naively 
go in  closest analogy to the one presented above, there still is a 
potential of locating the origin of differences. The standard arguments for 
factorization 
based on $1/N_c$ counting rules are not necessarily applicable for the decays
involving heavy baryons, in particular when the velocities of the final
particles are small. This theoretical possibility must be carefully explored.
On the other hand, a detailed experimental information about $\Lambda_b$
hadronic width and, first of all, the value of $n_c$, can provide us
with an invaluable direct input.
\vspace*{.3cm}

{\bf Acknowledgements}: \hspace{0.2em} The authors
would like to thank G. Altarelli, P. Ball, I.~Bigi and J.~Rosner for
useful discussions, T. Browder
for  comments on the experimental situation.
One of the authors (B.B.) thanks I. Dunietz for explaining  his
results.
This work was supported in part by DOE under the grant number
DE-FG02-94ER40823, by NSF under the grant
number PHY 92-13313 and by the
Israel Academy of Sciences and the VPR Technion fund.

\end{document}